\documentclass[aip,jmp,12pt,showkeys,amsmath]{revtex4-1}
\pdfoutput=1 % if your are submitting a pdflatex (i.e. if you have
\usepackage{bbold}
\usepackage[dvipsnames]{xcolor}
\newcommand{\be}{\begin{equation}}
\newcommand{\ee}{\end{equation}}
\newcommand{\ea}{\end{eqnarray}}
\newcommand{\ba}{\begin{eqnarray}}
\def\I{\mathbb{1}}

\newcommand{\wt}{\widetilde}

\newcommand{\Tr}{\operatorname{Tr}}
\newcommand{\pf}{\operatorname{pf}}
\def\dotwtxi#1{\dot{\widetilde\xi^{#1}\!\!\!}\,\,\,}
\begin{document} 
\title{\boldmath Noncommutative Mapping from the symplectic formalism}
\author{M. A. De Andrade }
\email{marco@fat.uerj.br}
\affiliation{Departamento de Matem\'{a}tica, F\'{\i}sica e Computa\c{c}\~{a}o, Faculdade de Tecnologia, \\ Universidade do Estado do Rio de Janeiro,\\
Rodovia Presidente Dutra, Km 298, P\'{o}lo Industrial, CEP 27537-000, Resende-RJ, Brazil.}
\author{C. Neves}
\email{clifford@fat.uerj.br}
\affiliation{Departamento de Matem\'{a}tica, F\'{\i}sica e Computa\c{c}\~{a}o, Faculdade de Tecnologia, \\ Universidade do Estado do Rio de Janeiro,\\
Rodovia Presidente Dutra, Km 298, P\'{o}lo Industrial, CEP 27537-000, Resende-RJ, Brazil.}
%
%\date{}

\begin{abstract}
The Bopp's shifts will be generalized through symplectic formalism. A special procedure, like a ``diagonalization", which drives the completely deformed symplectic matrix to the standard symplectic form was found as suggested by Faddeev-Jackiw. Consequently, the correspondent transformation matrix guides the mapping from commutative to noncommutative (NC) phase-space coordinates. The Bopp's shifts may be directly generalized from this mapping. In this context, all the NC and scale parameters, introduced into the brackets, will be lifted to the Hamiltonian. Well known results, obtained using $\star$-product, will be reproduced without to consider that the NC parameters are small$(<<1)$. Besides, it will be shown that different choices for NC algebra among the symplectic variables generates distinct dynamical systems, which they may not even connect with each other, and that some of them can preserve, break or restore the symmetry of the system. Further, we will also discuss the charge and mass rescaling in a simple model.

\end{abstract}
%
%\pacs{}
%
\keywords{Noncommutative Geometry, Symplectic Manifold, Quantum Mechanics}
\maketitle
\flushbottom
\section{Introduction}
\label{sec:intro}
In this last few decades, NC theories have been extensively studied. This happens because some results in string theory\cite{ST,STA} suggest that space-time coordinates may not commute\cite{snyder}, since it seems to be relevant to the quantization of D-branes in  background magnetic $(B_{\mu\nu})$ fields\cite{ST1,ST1A} and also because NC space-time coordinates is an alternative mechanism for Lorentz-invariance breaking\cite{JC,JC1,JC2}. Despite of great interest today, the noncommutativity of coordinates is an older idea presented at the beginning of the quantum theory, indeed, in some assumptions on its \textit{quantized} differential geometry\cite{dirac,dirac1} or in the description of nonrelativistic electrons of mass \textit{m} on a plane subject to a strong perpendicular magnetic field \textit{B} in the lowest Landau level\cite{peierls}. However, ever since Heisenberg's paper\cite{heisenberg1,heisenberg2} on uncertainty principle, there has been wide hesitancy in consider, simultaneously, classical-valued positions and momenta variables in any meaningful formula expressing quantum behavior, since these are incompatible observables. However, H. Groenewold\cite{HG} and J. Moyal\cite{MOYAL} provided an original technical solution to the problem above, \textit{i.e.}, they developed a special binary operation, the $\ast$-product, which it preserves the classical nature of positions$(q_i)$ and momenta$(p_i)$, but it also permits $q_i$ and $p_i$ to combine in a way  that it is equivalent to the familiar operator algebra of Hilbert space quantum theory. This original technical solution bases on the introduction of a Wigner phase space distribution function\cite{wigner} with deformed product and brackets (it is not the Poisson brackets), whose they correspond to the quantum commutators brackets. Due to this, essential aspects of quantum mechanics can be given a classical formalism in terms of the $\ast$-product. The deformation structure in the symplectic space\cite{flato} and the uniqueness of this formulation was systematically analyzed and mathematically consolidated\cite{fedosov} and, more recently,  it was proved that any finite-dimensional Poisson manifold can be canonically quantized in the sense of deformation quantization\cite{kont}. This alternative quantum framework is called \textit{Quantum Mechanics in Phase Space} (QMPS)\cite{QMPS1,QMPS2,QMPS3,QMPS4}. Inspired in these articles, we propose a formalization and generalization of the Bopp's shifts based on the NC symplectic induction procedure\cite{ANO2} which it begins with the deformation of the brackets among the fields, embraced by the symplectic variables, which it is identified as being a general and completely noncommutative deformed symplectic matrix (two-form tensor). This procedure changes the original canonical brackets among the variables, introducing $\,2n(n - 1)\,$ NC and $n$ scale parameters, which rescale the standard Poisson brackets. The transformation matrix among the newer symplectic variables (NC deformed system) and the older one (original system) is determined from the general and completely noncommutative deformed symplectic matrix.
The procedure to reduce a nonconstant deformed symplectic matrix into the standard one allows to obtain the standard Poisson brackets among the fields, which was point out by Faddeev and Jackiw\cite{FJ2}; here we obtain such ``\textit{symplecticzation}'' mechanism. With this mechanism, it is possible to render the NC geometry features into the dynamics, since the NC and scale parameters are lifted to the Hamiltonian and Lagrangian. This opens the possibility to investigate NC (\textit{quantum}) features at the commutative (\textit{classical}) level. Applying the noncommutative mapping into the $n$-dimensional dynamical system, the correspondent newer Lagrangian and Hamiltonian might to embrace, at the lower energy level, the NC (``\textit{quantum}'') features into them; at this point, the canonical quantized process can be applied, since the new brackets, now, might correspond to the quantum commutators. This paper is organized as follows: in Sec. \ref{sec:2}, the NC symplectic induction formalism will be presented, since it is the base for the mathematical formulation of the noncommutative mapping. In Sec. \ref{sec:3}, a simplified noncommutative mapping in a 4-dimensional dynamical system will be presented in order to allows a straightforward application in Sec. \ref{sec:4}. In Sec. \ref{sec:4}, the NC mapping will be applied in some systems in order to illustrate the procedure and also to explore new results. In Sec. \ref{sec:5}, conclusions will be presented.

\section{The Noncommutative Symplectic Induction Formalism}
\label{sec:2}
The symplectic formalism\cite{FJ2} is a powerful tool in the field theory and it was extended to also deal with constrained systems\cite{FJ3,FJ4,FJ5}, to induce symmetries into non-invariant systems\cite{ANO1,ANO1a,ANO1b,ANO1c}, and as well as to give an alternative way to introduce the Clebsch parameters\cite{WOCN} into some models. Further, it was also extended to induce noncommutativity into commutative systems\cite{ANO2,ANO2a,ANO2b}. There are a lot of ways to introduce NC\cite{MOYAL,mezincescu,ST1A,bopp1,bopp2,djemai1,djemai2,gosson2,gosson3,gosson1} into a system, however, a brief presentation of the NC symplectic induction formalism\cite{ANO2,ANO2a,ANO2b} will be necessary, since the Boop's shifts will be mathematically generalized through the symplectic framework.

\subsection{Setting the structure}
\label{building}

One will consider the systems whose dynamical equations can be derived from the general first-order Lagrangian
\be
\wt{L}(\wt{\xi},\dot{\wt\xi})=\wt{a}_{\alpha}(\wt\xi)\,\dotwtxi\alpha-H(\wt\xi),~~~ \alpha = 1,\ldots,{2n}, \label{NC-Lag-NCbasis}
\ee
where the overdot notation was employed for time derivative. Also, the configuration coordinate $\,\wt{q}_i\,$ and its conjugated momenta $\,\wt{p}_i\,$ were arranged in the 2$n$-component phase-space coordinate as 
\be
\wt\xi^\alpha=(\wt{q}_i\,,\,\wt{p}_i), ~~~  i = 1,\ldots,{n}.     \label{ph-sp-coord}
\ee
The arbitrary phase-space dependent functions $\,\wt{a}_{\alpha}(\wt\xi)\,$ and $\,H(\wt\xi)\,$ are respectively the FJ (Faddeev-Jackiw) coefficient and the Hamiltonian, the latter behaves as a potential function in a Lagrangian expressed in terms of the phase-space coordinate \cite{Jackiw}.
We will employ the following notations for derivatives: ${\partial}/{\partial \xi^\alpha}\equiv\partial_\alpha$\,;~ ${\partial}/{\partial {\wt\xi}^\alpha}\equiv{\wt\partial}_\alpha$.
Thus, the Euler-Lagrange equation of motion of the Lagrangian given in Eq.(\ref{NC-Lag-NCbasis}) can be obtained as
\be
{\wt{f}}_{\alpha\beta}\,\dotwtxi\beta={\wt\partial}_\alpha{H} ~,     \label{EqMovSymp}
\ee
where,
\be
\wt{f}_{\alpha\beta}\equiv\wt\partial_\alpha\,\wt{a}_\beta-\wt\partial_\beta\,\wt{a}_\alpha ~, \label{varsymb}
\ee
which it is a general antisymmetric symbol, depending of the phase-space coordinate.

On the other hand, let's consider the Lagrangian
\be
{L}({\xi},\dot{\xi})={a}_{\alpha}(\xi)\,{\dot\xi}^\alpha-H(\xi),~~~ \alpha = 1,\ldots,{2n}, \label{Lag-C}
\ee
whose equation of motion has the simplest antisymmetric constant symbol,
\be
{f}_{\alpha\beta}=\partial_\alpha\,{a}_\beta-\partial_\beta\,{a}_\alpha,  \label{consymb}
\ee
where $\,{f}_{\alpha\beta}$ are the elements of the \emph{inverse} ($f^{-1}$) of the usual $2n\times2n$ symplectic  matrix
\be
{f}=\left(\!\!
\begin{array}{rr}
  0 ~&~ \I  \cr
-\I ~&~  0
\end{array}             \label{symp}
\right)
\ee
with elements ${f}^{\alpha\beta}$. Then, the equation of motion of the Lagrangian given in Eq.(\ref{Lag-C}) may be read as
\be
{\dot\xi}^\alpha={f}^{\alpha\beta}\,{\partial}_\beta{H} ~,     \label{EqMovSymp2}
\ee
and its components $\,{\dot\xi}^\alpha=(\dot{q}_i\,,\,\dot{p}_i)$ correspond to the usual Hamilton equations of motion. 
For this constant ${f}_{\alpha\beta}$, the FJ coefficient ${a}_\alpha$ gets the linear relation,
\be
{a}_\alpha(\xi)=\frac12\,\xi^\beta\,{f}_{\beta\alpha}~,     \label{kincoeff}
\ee
and the Lagrangian given in Eq.(\ref{Lag-C}) takes the usual shape, which in terms of the phase-space coordinate, disregarding a total time derivative inexpressive in the action of any Lagrangian, it may be written as
\be
L(q,\dot{q})=p_i\,\dot{q}_i-H(p,q).           \label{L_config}
\ee

The Poisson bracket is one of the possibles starting point to quantize a theory, since the quantum constant $\hbar$ can be introduced through direct replacement of the Poisson bracket by the corresponding commutator: For two classical quantities $\,{F}(q,p)\,$ and $\,G(q,p)\,$ this replacement is represented by
\be
\label{CQ}
\{{F}\,,\,{G}\}\longrightarrow\frac{~1~}{~i\,\hbar~}\,\big[\widehat{F}\,,\,\widehat{G}\big]  \,,
\ee
where $\widehat{F}$ and $\widehat{G}$ are the corresponding quantum operators.
We can express the Poisson bracket of $\,{F}(\xi)\,$ and $\,{G}(\xi)\,$ through symplectic formalism with the help of the elements of the symplectic matrix given in Eq.(\ref{symp}) as
\be
\{{F}\,,\,{G}\}=\partial_\alpha{F}\,{f}^{\alpha\beta}\,\partial_\beta{G}  ~. \label{PB_geral}
\ee
From Eq.(\ref{PB_geral}), we can calculate the Poisson brackets of the phase-space coordinates directly as
\be
 \{\xi^\alpha\,,\,\xi^\beta\}={f}^{\alpha\beta} ~. \label{PB_basic}
\ee
Putting back this term into the Eq.(\ref{PB_geral}), we get
\be
\{{F}\,,\,{G}\}=\partial_\alpha{F}\,\{\xi^\alpha\,,\,\xi^\beta\}\,\partial_\beta{G} ~. \label{PB_geral2}
\ee

\subsection{Adding noncommutative and scale parameters into the system}

One strategy that can be followed to incorporate noncommutativity in the subject is taking Eq.(\ref{PB_basic}) as a starting point, realizing that the elements of ${f}$ can be sorted by {\em direct} Poisson brackets (with ${i}={j}$) which correspond to the canonical Poisson brackets, and {\em crossed} Poisson brackets (with ${i}\neq{j}$) , the latter are given by
\be
\{{q}_i\,,\,{q}_j\}=0~,~~~~\{{q}_i\,,\,{p}_j\}=0~,~~~~\{{p}_i\,,\,{q}_j\}=0~,~~~~\{{p}_i\,,\,{p}_j\}=0~,~~~~{i}\neq{j}~.      \label{PB_crossed}
\ee
Then, we can add NC parameters into the system through a suitable change of basis $~\xi^\alpha \rightarrow \wt\xi^\alpha~$ ($\text{commutative basis}\rightarrow\text{noncommutative basis}$). So, in order to accomplish this plan, we just need {\em rescale} the {\em direct} Poisson brackets and {\em replace} the zero elements corresponding to the {\em crossed} Poisson brackets by NC parameters according to the problem to be investigated. In the most general way, one must consider that the Poisson brackets on noncommutative basis may be mapped via the following equation,
\be
\{\wt\xi^\alpha\,,\,\wt\xi^\beta\}=\wt{f}^{\alpha\beta} ~,    \label{PB_news2}
\ee 
where $\wt{f}^{\alpha\beta}$ are the elements of the full $2n\times2n$ antisymmetric matrix $\wt{f}$ whose elements depend of the phase-space coordinate. The matrix $\wt{f}$ will be the general deformed symplectic matrix on the noncommutative basis.
In order to facilitate the presentation, let's briefly restrict to the $2$-dimensional configuration space and also to employ a new representation for the symplectic matrix where the rows and columns corresponding to $q_2$, $p_1$ are swapped. In this new representation, the four-component phase-space coordinate is arranged as $\,\xi^\alpha=({q}_1\,,\,{p}_1\,,\,{q}_2\,,\,{p}_2)$ and the $4\times4$  symplectic matrix  gets the block-diagonal form
\be\arraycolsep=1.4pt\def\arraystretch{1.1}
{f}=\left(
\begin{array}{cc}
\varepsilon ~&~  0           \cr
 0          ~&~  \varepsilon  
\end{array}                          \label{symp2}
\right),
\ee
where $\varepsilon$ is the 2$\times$2 antisymmetric matrix whose elements are the two-dimensional Levi-Civita symbol with $\varepsilon_{12}=1$. So, on the commutative basis, the direct Poisson brackets correspond to the elements of the 2$\times$2 $\varepsilon$-matrices in Eq.(\ref{symp2}) and the crossed Poisson brackets correspond to the elements of the 2$\times$2 null-matrices; on the noncommutative basis, the direct Poisson brackets will be rescaled and the crossed Poisson brackets will be given by
\be
\{\wt{q}_1\,,\,\wt{q}_2\}=\theta_{\bar{1}\bar{1}}~,~~~~\{\wt{q}_1\,,\,\wt{p}_2\}=\theta_{\bar{1}\bar{2}}~,~~~~\{\wt{p}_1\,,\,\wt{q}_2\}=\theta_{\bar{2}\bar{1}}~,~~~~\{\wt{p}_1\,,\,\wt{p}_2\}=\theta_{\bar{2}\bar{2}} ~,    \label{PB_news}
\ee
and they can be arranged in a full antisymmetric $4\times4$ matrix $\wt{f}$ as
\be\arraycolsep=1.4pt\def\arraystretch{1.2}
\wt{f}=\left(
\begin{array}{cc}
g_1\,\varepsilon  ~&~  {\Theta} \cr
-{\Theta}^T       ~&~  g_2\,\varepsilon
\end{array}\right) ~,                  \label{NC-symp2}
\ee
where $\,g_1$, $g_2\,$ are arbitrary scale parameters and ${\Theta}$ is the 2$\times$2 matrix whose elements are the arbitrary NC parameters $\theta_{\bar\imath\bar\jmath}$. 

Let's restrict to the case in which all the scale and NC parameters of the Poisson brackets of the Eq.(\ref{PB_news2}) are integrally transferred to the Lagrangian of Eq.(\ref{NC-Lag-NCbasis}). To guarantee that this happens, we may verify that $\det(\wt{f})=[g_1\,g_2-\det({\Theta})]^2$. Then, ensuring that $\det({\Theta})\neq{g_1\,g_2}$, we can proceed working with $\wt{f}^{-1}$, with elements $\,\wt{f}_{\alpha\beta}$, and readily to provide through the Eq.(\ref{varsymb}) all the scale and NC parameters for the $\wt{a}_{\alpha}(\wt\xi)$ on the first term of the Lagrangian given in Eq.(\ref{NC-Lag-NCbasis}). For $n$-dimensional configuration space, the general deformed symplectic $2n\times{2n}$ matrix $\wt{f}$ will be embrace $2n(n-1)$ NC and $n$ scale parameters. The generalization to higher configuration space is straightforward and it may be obtained from the results presented in the Appendix \ref{appendixA}.

After this, one may to follow the suggestion read in the Faddeev and Jackiw article\cite{FJ2} and to find a transformation matrix to change from the noncommutative basis to the commutative basis and then to express the henceforth restricted Lagrangian of the Eq.(\ref{NC-Lag-NCbasis}) in terms of the latter basis where the direct Poisson brackets are the canonical ones and the crossed Poisson brackets vanish. So, we can realize that an expression containing the transformation matrix can be readily obtained with the substitution ${F}=\wt\xi^\alpha$ and ${G}=\wt\xi^\beta$ in Eq.(\ref{PB_geral2}),
\be
\partial_\kappa{\wt\xi^\alpha}\,\{\xi^\kappa\,,\,\xi^\lambda\}\,\partial_\lambda\wt\xi^\beta = \{\wt\xi^\alpha\,,\,\wt\xi^\beta\} ~. \label{PBn-PBo}
\ee
The transformation matrix will be represented by ${R}$ such that,
\be
\partial_\beta\wt\xi^\alpha={R}^{\alpha}_{~\beta}  ~.   \label{transxi_res}
\ee
Then, taking into account Eq.(\ref{PB_basic}) and Eq.(\ref{PB_news2}), we can express Eq.(\ref{PBn-PBo}) in matrix form as
\be
{R}\,{f}\,{R}^T = \wt{f}  ~.                                \label{R_prop_1}
\ee
In order to get the matrix ${R}$ from the previous equation, first it's necessary to ensure that ${R}$ is invertible: It follows from Eq.(\ref{R_prop_1}) that ${[\det({R})]}^2=\det(\wt{f})$. Thus, to get ${R}$ invertible, we must request the same condition which was provided before for that $\wt{f}$ were invertible. After that, we can isolate ${f}$ by moving ${R}$ and ${R}^T$ to the other side of Eq.(\ref{R_prop_1}) and then to calculate the inverse of the resulting expression, which can be read as
\be
{R}^{T}\wt{f}^{-1}\,{R}={f}^{-1} \label{R_prop_2}.
\ee
The matrix product obtained \ by \ juxtaposing \ the \ expressions \ given in Eq.(\ref{R_prop_1}) and Eq.(\ref{R_prop_2}) in a way that its right side is given by $\wt{f}\,{f}^{-1}$ is read as
\be
{R}\,{f}{R}^T\,{R}^T\wt{f}^{-1}\,{R}=\wt{f}\,{f}^{-1}.
\ee
The previous equation can be worked to keep ${R}^T{R}^T$ isolated, and after a transposition it follows that
\be
{R}{R}=\wt{f}\left({R}^{-1}\wt{f}\,{f}^{-1}\,{R}^{-1}\right)^{T}{f}^{-1}.  \label{RR}
\ee
At this point, considering the  \textit{ansatz} $\,\wt{f}\,{f}^{-1}={R}{R}\,$ and inputting the right side of this expression in the middle of the term between parenthesis of the Eq.(\ref{RR}), we get as output the same expression, ${R}{R}=\wt{f}\,{f}^{-1}$. It's like if you asked someone: Is that? and then he answered: Yes that is. It follows that
\be
{R}=\sqrt{\wt{f}\,{f}^{-1}}.  \label{matrix-R}
\ee
 So, according Eq.(\ref{R_prop_2}), the matrix $R^{-1}=\sqrt{{f}\,\wt{f}^{-1}}$ effects the ``\textit{symplecticzation}''\cite{FJ2} of a full antisymmetric nonconstant $\wt{f}$ matrix into the standard form $f$ given in Eq.(\ref{symp}).

\subsection{Noncommutativity transformations}

From Eq.(\ref{transxi_res}) and Eq.(\ref{matrix-R}), one takes the NC transformations for the differential of the phase-space coordinate (contravariant) and for the derivative operator (covariant) respectively as
\ba
d\xi^{\alpha}     &\stackrel{\text{NC}}{\longrightarrow}& d\wt\xi^{\alpha}={R}^{\alpha}_{~\beta}\,d\xi^{\beta} ~, \label{contravector} \\
\partial_{\alpha}&\stackrel{\text{NC}}{\longrightarrow}& \wt\partial_{\alpha}={{R}^{-1}}^{\beta}_{~\alpha}\,\partial_{\beta} ~. \label{covector}
\ea
Eq.(\ref{R_prop_2})  in components is expressed as
\be
{R}^{\kappa}_{~\alpha}\;\wt{f}_{\kappa\lambda}\;{R}^{\lambda}_{~\beta}={f}_{\alpha\beta} \label{R_prop_2_comp}.
\ee
The substitution of Eq.(\ref{varsymb}) and Eq.(\ref{consymb}) in Eq.(\ref{R_prop_2_comp}) settles that the derivative of the FJ coefficient has the following NC transformation:

\be
({\wt\partial}_\alpha\,\wt{a}_\beta)={R^{-1}}^{\kappa}_{~\alpha}\;{R^{-1}}^{\lambda}_{~\beta}\;(\partial_\kappa\,{a}_\lambda)~. \label{cotensor}
\ee
Considering Eq.(\ref{contravector}) and Eq.(\ref{cotensor}), results from the differential identity $\,d\,{\wt{a}}_{\alpha}=({\wt\partial}_\beta\,\wt{a}_\alpha)\,d{\wt\xi}^\beta\,$ that $d{a}_{\alpha}$ behaves under NC transformation as a covariant vector
\be
d{a}_{\alpha} \stackrel{\text{NC}}{\longrightarrow} d\,{\wt{a}}_{\alpha}={{R}^{-1}}^{\beta}_{~\alpha}\,d{a}_{\beta} ~. \label{covector2}
\ee

\subsection{General symplectic matrix case}

Apart an additive exact differential inexpressive in the action of any Lagrangian one arrives at 
\be
\wt{a}_\alpha\,d{\wt\xi}^\alpha=-(d\,\wt{a}_\alpha)\,{\wt\xi}^\alpha 
\ee
After employ the covariant vector transformation given in Eq.(\ref{covector2}) we get
\be
\wt{a}_\alpha\,d{\wt\xi}^\alpha=-(d{a}_\beta)\;{{R}^{-1}}^{\beta}_{~\alpha}\,{\wt\xi}^\alpha~.
\ee
Disregarding again an additive exact differential, one can read
\ba
\wt{a}_\alpha\,d{\wt\xi}^\alpha
&=&{a}_\beta\;d\big({{R}^{-1}}^{\beta}_{~\alpha}\,{\wt\xi}^\alpha\big)  \nonumber \\
&=&{a}_\beta\;d\big({{R}^{-1}}^{\beta}_{~\alpha}\big)\,{\wt\xi}^\alpha+{a}_\beta\;d\xi^\beta~.
\ea
The first term on the right hand side of the previous equation breaks the NC invariance of the FJ one-form, Thus the NC Lagrangian given in Eq.(\ref{NC-Lag-NCbasis}) can be read as
\be
\wt{L}(\xi,\dot\xi)={a}_\beta\,{\dot\xi}^\beta-{H}(\wt\xi)+{a}_\beta\,{\dot{R}^{-1}{}}^{\beta}_{~\alpha}\;{\wt\xi}^\alpha~.
\ee
Finally, from Eq.(\ref{kincoeff}), we get
\be
\wt{L}(\xi,\dot\xi)=\frac12\,\xi^\alpha\,{f}_{\alpha\beta}\,{\dot\xi}^\beta-{H}(\wt\xi)+\frac12\,\xi^\alpha\,{f}_{\alpha\beta}\,{\dot{R}^{-1}{}}^{\beta}_{~\kappa}\;{\wt\xi}^\kappa~,                                  \label{NC-Lag-Cbasis}
\ee
whose first term, disregarding a total derivative, can be written in the usual $\,p_i\,\dot{q}_i\,$ way and $\,\wt\xi=\int\!R\,d\xi$. The extra term,
\be
\delta{\wt{L}} = \frac12\,\xi^\alpha\,{f}_{\alpha\beta}\,{\dot{R}^{-1}{}}^{\beta}_{~\kappa}\;{\wt\xi}^\kappa~,
\ee
it can disappear when some choices for NC algebra among the variables are considered. From this relation, a suitable change of basis $~\xi^\alpha \rightarrow \wt\xi^\alpha~$ ($\text{commutative basis}\rightarrow\text{noncommutative basis}$) can preserve, break or restore a symmetry of the system, now without the necessity to extend the phase-space with the introduction of the Wess-Zumino fields.

\subsection{Constant symplectic matrix case}

In a {\em very special case} where ${R^{-1}}^{\beta}_{~\alpha}$ in Eq.(\ref{covector2}) do not depend of the phase-space coordinate, we get after an integration that  the NC transformation for the FJ coefficient will be given by
\be
\wt{a}_\alpha={R^{-1}}^{\beta}_{~\alpha}\,{a}_\beta ~,        \label{co_kin_coeff}
\ee
which is a covariant transformation. From the differential relation given in Eq.(\ref{contravector}), we get a linear NC transformation for the phase-space velocity
\be
\dotwtxi{\alpha}={R}^{\alpha}_{~\beta}\,{\dot\xi}^{\beta} ~,    \label{contra_sym_vel}
\ee
which is a contravariant transformation. So, with Eq.(\ref{co_kin_coeff}) and Eq.(\ref{contra_sym_vel}), we get the following NC invariant
\be
\wt{a}_\alpha\,\dotwtxi\alpha={a}_\alpha\,{\dot\xi}^\alpha ~,  \label{kin-inv}
\ee
which corresponds to the NC invariance of the first term of the Lagrangian from Eq.(\ref{NC-Lag-NCbasis}). Using Eq.(\ref{kincoeff}), and considering that $\,{H}(\wt\xi)={H}({R}\xi)={\wt{H}}(\xi)$, it can be read as
\be
\wt{L}(\xi,\dot\xi)=\frac12\,\xi^\alpha\,{f}_{\alpha\beta}\,{\dot\xi}^\beta-{\wt{H}}(\xi).     \label{NC-Lag-Cbasis-const}
\ee
The equation of motion from the Lagrangian given in previous equation is expressed by
\be
\dot\xi^\alpha={f}^{\alpha\beta}\,\partial_\beta{\wt{H}}  ~, \label{EqMov_NC}
\ee
whose components are the Hamilton's equations of motion, that now carry the NC parameters. 
The Lagrangian given in Eq.(\ref{NC-Lag-Cbasis-const}) can be directly expressed in the configuration space, disregarding a total derivative term, as
\be
\wt{L}(q,\dot{q})=p_i\,\dot{q}_i-\wt{H}(p,q)~. \label{NC-Lag-config}
\ee
So, in this simplest case, with the change of the system prescription from the noncommutative basis to the commutative basis, one will find that all the NC-ingredients were transferred to the Hamiltonian and they could be interpreted as being an external (unknown) potential, or a background field, or Lorentz symmetry breaking mechanism or even a mass generation mechanism.

\section{Representation for two-dimensional space with {$4\times4$} matrices}
\label{sec:3}

Consider a two-dimensional space, whose configuration coordinates ${q}_i$ and their conjugated momenta ${p}_i$ are arranged in the four-component phase-space coordinate as $\xi^\alpha=(q_1, p_1, q_2, p_2)$. One can express the symplectic matrix in noncommutative basis through blocks as
\be
\wt{f}=\left(
\begin{array}{cc}
{g}_1\,\varepsilon ~&~  \Theta \cr
-\Theta^T          ~&~   {g}_2\,\varepsilon
\end{array}\right) ~ , \label{noncommut_paired}
\ee
where ${g}_1$, ${g}_2$ are scale parameters; the $2\times2$ antisymmetric matrix $\varepsilon$ has as elements the Levi-Civita two-dimensional symbol with $\varepsilon_{12}=1$; and the $2\times2$ matrix
\be
{\Theta}=
\left(
\begin{array}{rl}
\{\wt{q}_1,\wt{q}_2\} ~&~  \{\wt{q}_1,\wt{p}_2\}  \cr
\{\wt{p}_1,\wt{q}_2\} ~&~  \{\wt{p}_1,\wt{p}_2\}
\end{array}\right)  ~ \label{Theta}
\ee
has as elements the NC parameters identified as the Poisson brackets.
One may proceed swapping the second and third rows as well as the columns of $\wt{f}$ in the Eq.(\ref{noncommut_paired}) for correspond to the phase-space coordinate in the conventional order, $\xi^\alpha=(q_1,q_2,p_1,p_2)$. Then, by means of a couple of new $2\times2$ matrices containing two parameters each one,
\be
{\Sigma}= 
\left(
\begin{array}{cc}
0                       ~&~  \{\wt{q}_1,\wt{p}_2\} \cr
-\{\wt{p}_1,\wt{q}_2\}  ~&~             0
\end{array}\right)                                     \label{CPB_SR_qp}
~~~,~~~
{D}_g= 
\left(
\begin{array}{cc}
{g}_1  ~&~  0 \cr
0      ~&~  {g}_2
\end{array}\right)~,
\ee
and the remaining NC parameters, the symplectic matrix and its inverse in noncommutative basis can be read as
\be\arraycolsep=1.4pt\def\arraystretch{1.5}
\wt{f}=\left(
\begin{array}{cc}
\{\wt{q}_1,\wt{q}_2\}\,\varepsilon       ~~&~  {D}_{g}+{\Sigma} \cr
-{D}_{g}-{\Sigma}^T                      ~~&~  \{\wt{p}_1,\wt{p}_2\}\,\varepsilon
\end{array}\right) ~,
\ee
\be\arraycolsep=1.4pt\def\arraystretch{1.5}
\wt{f}^{-1}=\frac{1}{{g}_1\,{g}_2-{b}^2}\left(
\begin{array}{cc}
\{\wt{p}_1,\wt{p}_2\}\,\varepsilon          ~~&~~ \varepsilon\,{D}_{g}\,\varepsilon+{\Sigma}^T \cr
-\varepsilon\,{D}_{g}\,\varepsilon-{\Sigma} ~~&~~  \{\wt{q}_1,\wt{q}_2\}\,\varepsilon
\end{array}\right)~,   \label{noncommut-stand}
\ee 
where ${b}^2=\det(\Theta)$. The matrix that perform transformations from commutative to noncommutative basis, ${R}$ defined in Eq.(\ref{matrix-R}), and its inverse will be given by
\be\arraycolsep=1.4pt\def\arraystretch{1.5}
{R}=
\left(
\begin{array}{cc}
{D}_{s}+\frac{1}{{s}_1+{s}_2}\,{\Sigma}                        ~~&~~ -\frac{1}{{s}_1+{s}_2}\,\{\wt{q}_1,\wt{q}_2\}\,\varepsilon\cr
\frac{1}{{s}_1+{s}_2}\,\{\wt{p}_1,\wt{p}_2\}\,\varepsilon      ~~&~~ {D}_{s}+\frac{1}{{s}_1+{s}_2}\,{\Sigma}^T
\end{array}\right)~,  \label{R-stand}
\ee
\be\arraycolsep=1.4pt\def\arraystretch{1.5}
{R}^{-1}=\frac{1}{\sqrt{{g}_1\,{g}_2-{b}^2}}
\left(
\begin{array}{cc}
-\varepsilon\,{D}_{s}\,\varepsilon-\frac{1}{{s}_1+{s}_2}\,{\Sigma} ~~&~~  \frac{1}{{s}_1+{s}_2}\,\{\wt{q}_1,\wt{q}_2\}\,\varepsilon\cr
-\frac{1}{{s}_1+{s}_2}\,\{\wt{p}_1,\wt{p}_2\}\,\varepsilon     ~~&~~ -\varepsilon\,{D}_{s}\,\varepsilon-\frac{1}{{s}_1+{s}_2}\,{\Sigma}^T
\end{array}\right),          \label{R-stand_app}
\ee
where ${D}_{s}$ is the auxiliary diagonal $2\times2$ matrix
\be
{D}_{s}= 
\left(
\begin{array}{cc}
{s}_1  ~&~  0 \cr
0      ~&~ {s}_2
\end{array}\right)~,
\ee 
with
\be
{s}_i^2= {g}_i+\frac{2\,{b}^2\,\sqrt{{g}_1\,{g}_2-{b}^2}-{b}^2\,({g}_1+{g}_2)}{({g}_1-{g}_2)^2+4\,{b}^2}~.    \label{s_parameter}
\ee
Then, it follows some identities which can be useful for calculations:
\ba
&&{s}_i^2+\frac{{b}^2}{({s}_1+{s}_2)^2}={g}_i  ~; \\ 
&&{s}_1\,{s}_2-\frac{{b}^2}{({s}_1+{s}_2)^2}=\sqrt{{g}_1\,{g}_2-{b}^2} ~; \\
&&({s}_1+{s}_2)^2={g}_1+{g}_2+2\,\sqrt{{g}_1\,{g}_2-{b}^2}~; \\
&&{g}_i\,{s}_j-\frac{{b}^2}{{s}_1+{s}_2}={s}_i\,\sqrt{{g}_1\,{g}_2-{b}^2}~,~{j}\neq{i}\,; \\
&&{s}_i-\frac{{g}_i}{{s}_1+{s}_2}=\frac{\sqrt{{g}_1\,{g}_2-{b}^2}}{{s}_1+{s}_2}~.
\ea

\section{Applications}
\label{sec:4}

At this section, the NC mapping will be applied in order to introduce NC and scale parameters into a general model. This will be done in order to illustrate and shed some light on the question about the role played by the NC parameter into the model. It is important to notice that for any potential, written in terms of inverse power of $|\vec {x}|$, the NC induction approach could leads this potential to a very complex form, which it obstructs the straightforward calculation of the equation of motion and the implementation of any quantization process. However, the NC symplectic induction formalism and NC map give an alternative way to introduce the NC algebra into the model, even that the potential presents inverse power of $|\vec {x}|$.

\subsection{A particle in a general potential}
As the first example, we will consider a two-dimensional system where a particle, with mass \textit{m}, suffers the action of a general potential $V(q)$. This particle has its dynamic governed by the following Lagrangian density,
\be
\label{0270A}
{\cal{L}}=\frac {m \,\dot q_i^2}{2} - V(q), \texttt{ with } i=1,2.
\ee
In agreement with the NC symplectic induction formalism, this Lagrangian should be written in a first-order form as in Eq.(\ref{L_config}), then
\ba
\label{0270}
{\cal{L}}&=&p_i \, \dot q_i - \left[\frac{p_i^2}{2m} + V(q)\right],\nonumber\\
 &=&p_i \, \dot q_i - {\cal{H}},
\ea
where ${\cal{H}}=\frac {p_i^2}{2m} + V(q)$ is the Hamiltonian or, in the symplectic language, the symplectic potential. Now, we are ready to introduce the NC algebra into the model. To this end, we will propose the following NC brackets among the variables,
\ba
\label{0300}
\left\{\wt{q}_i,\wt{q}_j\right\}&=& \varepsilon_{ij}\theta,\nonumber\\
\left\{\wt{p}_i,\wt{p}_j\right\}&=& 0,\\
\left\{\wt{q}_i,\wt{p}_j\right\}&=&\delta_{ij}.\nonumber
\ea
Note that $\theta$ has canonical dimensions of $(\mathit{length})^2$ and, consequently, it introduces an ultra-violet scale to the problem if it is taken to be small. The transformation of phase-space coordinates, the commutative to the noncommutative one is implemented by
\be
\label{0404}
\wt\xi^\alpha=R^{\alpha}_{~\beta}\,\xi^\beta~. 
\ee
Considering the results given in Sec. \ref{sec:3}, we can follow the recipe given in Eq.(\ref{R-stand}) and to build the matrix $R$ taking 
${D}_{s}=\I$, $\Sigma=0$, $\{\wt{q}_1,\wt{q}_2\}=\theta$ and $\,\{\wt{p}_1,\wt{p}_2\}=0$. Then, using Eq.(\ref{0404}), we get the transformations that lead us to the NC Lagrangian, which in the shape of Eq.(\ref{NC-Lag-config}), it's given by
\be
\label{0371}
\wt{\cal{L}} =  p_i \, \dot q_i - \left[\frac{p_i^2}{2m}+V(q_1-\frac{\theta}{2}{p}_2\,,\,q_2+\frac{\theta}{2}{p}_1)\right].
\ee
This is the same result obtained when the $\ast$-product is implemented by Mezincescu.\cite{mezincescu} If the Coulomb potential is chosen, we can reproduce the result obtained by Chaichian \textit{et al.} \cite{chaichian}

After that, we will explore another possibility. To this end, the following NC brackets among the variables will be proposed,
\ba
\label{0410}
\left\{\wt{q}_i,\wt{q}_j\right\}&=& 0 ,\nonumber\\
\left\{\wt{p}_i,\wt{p}_j\right\}&=& \varepsilon_{ij}\theta,\\
\left\{\wt{q}_i,\wt{p}_j\right\}&=&\delta_{ij}.\nonumber
\ea
Now, the NC parameter $(\theta)$ has canonical dimensions of $(\mathit{mass})^2$ and, consequently, it introduces an infra-red scale if it is taken to be small. To reflect the NC brackets given in Eq.(\ref{0410}), we must set up in Eq.(\ref{R-stand}): 
${D}_{s}=\I$, $\Sigma=0$, $\{\wt{q}_1,\wt{q}_2\}=0$ and $\,\{\wt{p}_1,\wt{p}_2\}=\theta$. 
Thus, repeating the receipt used before, the NC first-order Lagrangian is obtained as 
\be
\label{0440}
\wt{\cal{L}} = p_i \, \dot q_i - \wt{\cal{H}}
\ee
where the NC Hamiltonian is
\ba
\label{0445}
\wt{\cal{H}}&=& \frac{1}{2\,m}\left[\left(p_1+\frac{\theta}{2}q_2\right)^2+\left(p_2-\frac{\theta}{2}q_1\right)^2\right] +V(q),\nonumber\\
\mbox{}&=& \frac{p_i^2}{2m}+ \frac{\theta}{2m}(p_i\varepsilon_{ij}q_j)+\frac{\theta^2}{8m}q_i^2+V(q)~.
\ea
From the Euler-Lagrange equation of motion for the canonical momenta, we get,
\ba
p_1 &=& m\, \dot q_1 - \frac{\theta}{2} q_2,\nonumber\\
p_2 &=& m\, \dot q_2 + \frac{\theta}{2} q_1.
\ea
Introducing these canonical momenta into the NC first-order Lagrangian given in Eq.(\ref{0440}), we get
\be
\label{0450}
\wt{\cal{L}} = \frac{m\,\dot q_i^2}{2} + \frac{\theta}{2}\,q_i\varepsilon_{ij}\dot q_j - V(q).
\ee
Therefore, distinct choices for the NC algebra among the brackets render distinct dynamic systems.

\subsection{Charged harmonic oscillator}
Assuming a system with the potential as being the two-dimensional harmonic oscillator, namely,
\be
\label{0460}
V(q) = \frac{m\omega^2}{2}(q_i^2),
\ee
where $\omega$ is the frequency, the NC first-order Lagrangian given in Eq.(\ref{0450}), renders to
\ba
\label{0465}
\wt{\cal{L}} &=& \frac{m\dot q_i^2}{2} + \frac{\theta}{2}q_i\varepsilon_{ij}\dot q_j - \frac{m\omega^2}{2}(q_i^2),\nonumber\\
&=& \frac{m\dot q_i^2}{2} + \frac{\theta}{2}q_i\varepsilon_{ij}\dot q_j - \frac{\theta^2}{8m}q_i^2 - \frac{m\omega^+\cdot\omega^-}{2}(q_i^2),
\ea
with
\be
\label{0466}
\omega^\pm=\omega\pm\frac{\theta}{2m}.
\ee
Writing the Lagrangian above in a first-order form, namely,
\be
\wt{\cal{L}} = p_i\cdot \dot{q}_i - \left(\frac{p_i^2}{2m} + \frac{\theta (p_i\varepsilon_{ij}q_j)}{2m} + \frac{\theta^2q_i^2}{4m} + \frac{m\omega^+\cdot\omega^-}{2}(q_i^2)\right),
\ee
the correspondent Hamiltonian is identified as being
\be
\wt{\cal{H}} = \frac{p_i^2}{2m} + \frac{\theta (p_i\varepsilon_{ij}q_j)}{2m} + \frac{\theta^2q_i^2}{4m} + \frac{m\omega^+\cdot\omega^-}{2}(q_i^2),
\ee
or
\be
\label{0466a}
\wt{\cal{H}} = \frac{p_i^2}{2m} + \frac{\theta (p_i\varepsilon_{ij}q_j)}{2m} + \frac{\theta^2q_i^2}{8m} + \frac{m\omega^2}{2}(q_i^2),
\ee
since the relation given in Eq.(\ref{0466}) was used. 

From the investigation done by Banerjee and Ghosh\cite{BG}, the Hamiltonian of a charged harmonic oscillator in an axially symmetric magnetic field is
\be
\label{0470}
{\cal{H}}=\frac{p_i^2}{2m}+\frac{e\cdot B(t)}{2m\cdot c}(p_i\varepsilon_{ij}q_j) + \frac{e^2\cdot {B}^2(t)}{8m\cdot c^2}q_i^2 +\frac{m\omega^2}{2}(q_i^2).
\ee
Comparing the Hamiltonian above with the one given in Eq.(\ref{0466a}), we obtain the following identity,
\be
\label{0480}
\theta = \frac{e\cdot B(t)}{c}.
\ee
Therefore, the NC version of two-dimensional harmonic oscillator represents an electron in a very simple Bohr model of the hydrogen atom, where there is a background magnetic field. Due to this, the motion of the electron can be split into components parallel and perpendicular to the magnetic field. In this setup, the NC parameter or the background magnetic field splits the original quantum level into three-levels, where one of the frequency remains unchanged and the others two frequencies changed to
\ba
\label{0490}
\omega^\pm&=& \omega \pm \frac{\theta}{2m},\nonumber\\
&=& \omega \pm \frac{e\cdot B(t)}{2m\cdot c}.
\ea
The Hamiltonian above has a structure very similar to the model of a charged particle in a specified electromagnetic field discussed in these papers\cite{Lewis1,Lewis2}, where the eigenstates are constructed of the invariant operator. In accordance with Banerjee and Ghosh\cite{BG}, the standard Zeeman level is given by
\ba
\label{0500}
E^{\pm}_n &=& (n+\frac 12)\hbar\omega \pm \left[ n+ (j+1/2)\right] \frac{\theta}{m},\nonumber\\
&=& (n+\frac 12)\hbar\omega \pm \left[ n+ (j+1/2)\right] \frac{e\cdot B}{m\cdot c},
\ea
where $B$ is a constant. With this approach, the presence of NC parameter into the system might be interpreted as being the origin of charge property of the particle and the interaction of this one with a background magnetic field. As expected, this changes the energy spectrum of the model, which it was shown above.

\subsection{Two independent harmonic oscillator}

In this section, we will explore new features from the NC approach discussed here. In order to do this, we will consider a system with two particles where each one presents mass \textit{m}. This system has its dynamics governed by the following Lagrangian density,
\be
\label{0510}
{\cal{L}}=\frac {m \, (\dot q_{1,i}^2+ \dot q_{2,i}^2)}{2} - V(q_{1,i},q_{2,i}), \texttt{ with } i=1,2.
\ee
The potential above is, for a while, a general interactive potential. The correspondent Hamiltonian is
\be
\label{0520}
{\cal{H}}=\frac {(p_{1,i}^2+ p_{2,i}^2)}{2m} + V(q_{1,i},q_{2,i}), \texttt{ with } i=1,2.
\ee
At this point, we started with the Lagrangian density given in Eq.(\ref{0510}), and propose a different NC brackets among the variables, namely,
\ba
\label{0720}
\left\{\wt{q}_{a,i},\wt{q}_{b,j}\right\}&=& 0 ,\nonumber\\
\left\{\wt{p}_{a,i},\wt{p}_{b,j}\right\}&=& \varepsilon_{ij}\cdot \delta_{ab}\cdot(-1)^{\delta_{a1}}\cdot\theta,\\
\left\{\wt{q}_{a,i},\wt{p}_{b,j}\right\}&=& \delta_{ij}\cdot \delta_{ab}.\nonumber
\ea
In this case we must set up in Eq.(\ref{R-stand}): 
${D}_{s}=\I$, $\Sigma=0$, $\{\wt{q}_1,\wt{q}_2\}=0$ and $\,\{\wt{p}_1,\wt{p}_2\}=\delta_{ab}\cdot(-1)^{\delta_{a1}}\cdot\theta\,$ 
to get the transformations that lead us to the NC Lagrangian
\be
\label{0745}
\wt{\cal{L}}= {p}_{a,i} \, \dot{{q}}_{a,i} - \wt{\cal{H}},
\ee
where the NC Hamiltonian is
\ba
\label{0748}
\wt{\cal{H}}&=& \frac{1}{2m}\left[\left({p}_{1,i}-\frac{\theta}{2}\varepsilon_{ij}{q}_{1,j}\right)^2+\left({p}_{2,i}+\frac{\theta}{2}\varepsilon_{ij}{q}_{2,j}\right)^2\right] + V(q_{a,i}) \nonumber\\
&=&\frac{p_{a,i}^2}{2m} +\frac{\theta}{2m}(-p_{1,i}\varepsilon_{ij}q_{1,j}+p_{2,i}\varepsilon_{ij}q_{2,j}) - \frac{\theta^2}{8m}q_{a,i}^2+V({q}_{a,i}).
\ea
The canonical momenta can be computed from Eq.(\ref{0745}),
\ba
\label{0800}
p_{1,i}&=& m\cdot \dot{q}_{1,i}+\frac{\theta}{2}\varepsilon_{ij}q_{1,j},\\
p_{2,i}&=& m\cdot \dot{q}_{2,i}-\frac{\theta}{2}\varepsilon_{ij}q_{2,j}.\nonumber
\ea
Introducing the canonical momenta given above into the first-order Lagrangian given in Eq.(\ref{0745}), we get
\be
\label{0810}
\wt{\cal{L}}= \frac{m}{2}\cdot (\dot{q}^2_{1,i}+\dot{q}^2_{2,i}) + \frac{\theta}{2}(\dot{q}_{1,i}\varepsilon_{ij}q_{1,j}-\dot{q}_{2,i}\varepsilon_{ij}q_{2,j})- V({q}_{a,j}).
\ee
Assuming that this model is a pair of an unitary charges of opposite sign in a magnetic field, where the NC parameter is interpreted as being the magnetic field, \textit{i.e.}, $\theta=B$, and doing an \textit{educated guess} for the potential,
\be
\label{0820}
V({q}_{a,j})=\frac{K}{2}(q_{1,j}-q_{2,j})^2 ,
\ee
the Lagrangian density given in (\ref{0810}) renders to
\be
\label{0830}
\wt{\cal{L}}= \frac{m}{2}\cdot (\dot{q}^2_{1,i}+\dot{q}^2_{2,i}) + \frac{B}{2}(\dot{q}_{1,i}\varepsilon_{ij}q_{1,j}-\dot{q}_{2,i}\varepsilon_{ij}q_{2,j})-\frac{K}{2}(q_{1,j}-q_{2,j})^2 ,
\ee
which it is the result discussed by Bigatti and Susskind\cite{BS} in Sec. I.A of this referred paper.

\subsection{Charge and mass rescaling}

The existence of the scale parameters on the NC scenario can have consequences on the mass values and charge values after the NC mapping. We verify these consequences on  the simple Coulomb model for the Hydrogen atom, turning on only the scales parameters in the NC algebra. To this end, we will propose the following NC brackets among the variables,
\ba
\label{0840}
\left\{\wt{q}_i,\wt{q}_j\right\}&=& 0,\nonumber\\
\left\{\wt{p}_i,\wt{p}_j\right\}&=& 0,\\
\left\{\wt{q}_i,\wt{p}_j\right\}&=&g\,\delta_{ij}, \ \texttt{ with } i=1,2,3. \nonumber
\ea
where $g$ is the scale parameter. The Coulomb model is described by the following Hamiltonian
\be
{\cal H}=\frac{p_{i}^2}{2\,m}-\frac{Z\,e^2}{\sqrt{q_{i}^2}}~.  \label{H_coulomb}
\ee
The symplectic matrices on the commutative and noncommutative basis for the phase-space coordinate in the conventional order ($\,q_{i}$\,,\,$p_{i}\,$) are respectively given by
\be
{f}=\left(\!\!
\begin{array}{cc}
 0           ~&~ \delta_{ij}  \cr
-\delta_{ij} ~&~  0
\end{array}             
\right)
~~~\text{and}~~~
\wt{f}=\left(\!\!
\begin{array}{cc}
 0              ~&~  g\,\delta_{ij}  \cr
-g\,\delta_{ij} ~&~  0
\end{array}             
\right).
\ee
The NC transformation matrix can be obtained from the symplectic matrices as
\be
{R}=\sqrt{\wt{f}\,{f}^{-1}}= 
\left(\!\!
\begin{array}{cc}
\sqrt{g}\,\delta_{ij} ~&~ 0 \cr
0                     ~&~   \sqrt{g}\,\delta_{ij}
\end{array}            
\right).
\ee
Finally, calcutating the NC phase-space coordinates on the commutative basis through the matrix $R$,  we obtain from Eq.(\ref{H_coulomb}) the NC Hamiltonian in the commutative framework with the rescaling of the mass and the charge,
\be
\wt{\cal H}=\frac{p_{i}^2}{2\,{m^\prime}}-\frac{Z\,{e^\prime}^2}{\sqrt{q_{i}^2}}~,
\ee
where 
\be
\label{massag}
m^\prime=\frac{m}{g} ~~~{\text{and}}~~~ {e^\prime}^2=\frac{e^2}{\sqrt{g}}~.
\ee
Thus, at the noncommutative scenario where, for instance, there are non-degenerescence in the mass spectrum and, consequently, we can adjust conveniently the values for the scale parameter to account for the diversity of the mass spectrum in the commutative scenario. Further, we can interpret the mass degenerescence presents in Eq.(\ref{massag}) in analogy with the mass spectrum arbitrariness obtained when the canonical quantization procedure is applied, which it arises when the quantum operator is ordered.

\section{Conclusion}
\label{sec:5}
In this article, the Bopp's shifts was generalized and systematized in the symplectic framework. Indeed, the Bopp's shifts was implemented by the transformation that maps the commutative and NC phase-space coordinate. Consequently, it reduces the NC deformed symplectic matrix to its canonical representation, as suggested by Faddeev and Jackiw\cite{FJ2}. Now, it is possible to introduce $\,2n(n-1)$ NC and $n$ scale parameters into a $n$-dimensional system in a wide, practical and easy way to setup different NC algebra. Consequently, it allows to explore the correspondent different contributions related to the noncommutativity. This result driven us to conclude that $\ast$-product was also generalized, since the usual $\star$-product induces simple Bopp's shifts\cite{bopp1,bopp2}: the NC mapping reproduces the results obtained when the $\star$-product is implemented without the necessity to constrain the NC parameters to be small$(<<1)$. Therefore, ultra-violet and infra-red divergence do not, necessarily, appear into the model. Furthermore, it was possible to investigate how gauge symmetry behavior can be related to the change of the basis $~\xi^\alpha \rightarrow \wt\xi^\alpha~$ ($\text{commutative basis}\rightarrow\text{noncommutative basis}$), \textit{i.e.}, the NC algebra among the variables induced into the system can preserve or break the previous gauge symmetry of the system. At this point, it is important to observe that it is possible to choose a NC algebra among the variables that induces a gauge symmetry into the system, \textit{i.e.}, now it is possible to transform systems with second-class constraints in first-class ones - gauge theories \cite{dirac} - without the necessity to extend the phase-space with the introduction of the auxiliary fields, as elegantly proposed by Batalin, Fradkin, Fradkina and Tyutin\cite{BFFT0,BFFT1,BFFT1a,BFFT2} and discussed in distinct framework\cite{ANO1,FS,CW}. In order to illustrate and put our procedure in a correct perspective with others works present in the literature, we apply the NC mapping in some very simple models: the one and two-dimensional harmonic oscillator. From this application some previous results presented in the literature were reproduced and it was also possible to show that distinct choices for the NC algebra among the symplectic variables generates distinct, in a dynamically point of view, NC systems that could be interpreted, at least in the lower energy level, as being \textit{quantum} versions of the correspondent commutative model. This was shown when it was considered to study the one harmonic oscillator, where the NC parameters in these system was responsible to split the quantum states of the hydrogen atom, as show by Banerjee and Ghosh.\cite{BG} Further, when the two harmonic oscillator was considered, it was possible to obtain the result obtained by Bigatti and Susskind\cite{BS} and, also, it was discussed and proposed an alternative interpretation for the scale parameters on the NC scenario: the mass and charge values depend on scale parameters, \textit{i.e.}, an arbitrariness in the mass spectrum arises, which it can explain the quantum features that arise when the NC system is mapped into its respective commutative one. This can be seen, in analogy with the quantum canonical procedure, as being the
mass spectrum arbitrariness due to the operator ordering ambiguity problem.
\begin{acknowledgments}
C. Neves and M. A. de Andrade would like to thank the Brazilian Research Agencies(CNPq and FAPERJ) for partial financial support.
\end{acknowledgments}
\appendix

\section{Representations for the symplectic and related matrices}
\label{appendixA}

\subsection{Representation for three-dimensional space with {$6\times6$} matrices}
\label{appendixA.1}

For a three-dimensional space whose configuration coordinates ${q}_i$ and their conjugated momenta ${p}_i$ are arranged in the six-component phase-space coordinate as
\be 
~\xi^\alpha=(q_1, p_1, q_2, p_2, q_3, p_3)~.
\ee
In order to embrace 12 (twelve) NC parameters in a compact manner, we will consider the three $2\times2$ matrices, ${\Theta}_{12},{\Theta}_{13} \text{ and }{\Theta}_{23}$, where each one embraces four independent NC parameters. Their respective determinants will be  represented as $b_{ij}^2$, corresponding to ${\Theta}_{ij}$. One will present only the $6\times6$ symplectic matrix $\wt{f}$ and its inverse in noncommutative basis:
\be\arraycolsep=1.4pt\def\arraystretch{1.3}
\wt{f}=\left(
\begin{array}{ccc}
g_1\,\varepsilon    ~&~  {\Theta}_{12}    ~&~  {\Theta}_{13} \cr
-{\Theta}_{12}^T  ~&~  g_2\,\varepsilon   ~&~  {\Theta}_{23} \cr
-{\Theta}_{13}^T  ~&~  -{\Theta}_{23}^T ~&~  g_3\,\varepsilon 
\end{array}\right).  
\ee
Let's consider the the symbol,
\be
b^2\equiv\left(g_1\,{b}_{23}^2+g_2\,{b}_{13}^2+g_3\,{b}_{12}^2\right)+\Tr\left(\Theta_{13}\,\varepsilon\,\Theta_{23}^T\,\varepsilon\,\Theta_{12}^T\,\varepsilon\right)~. 
\ee
One will highlight that the Pfaffian of the symplectic $6\times6$ matrix on noncommutative basis can be read as $\,\pf(\wt{f})=g_1\,g_2\,g_3-b^2$. Then, the inverse is given by
\be\arraycolsep=1.4pt\def\arraystretch{1.5}
\wt{f}^{-1}=\frac{1}{\pf(\wt{f})}\left(
\begin{array}{ccc}
       ({b}_{23}^2-g_2\,g_3)\varepsilon
~&~ -g_3\,\varepsilon{\Theta}_{12}\,\varepsilon+\varepsilon{\Theta}_{13}\varepsilon{\Theta}_{23}^T\,\varepsilon
~&~ -g_2\,\varepsilon{\Theta}_{13}\,\varepsilon-\varepsilon{\Theta}_{12}\varepsilon{\Theta}_{23}\,\varepsilon   \cr
       g_3\,\varepsilon{\Theta}_{12}^T\varepsilon+\varepsilon{\Theta}_{23}\varepsilon{\Theta}_{13}^T\,\varepsilon
~&~  ({b}_{13}^2-g_1\,g_3)\,\varepsilon
~&~ -g_1\,\varepsilon{\Theta}_{23}\,\varepsilon+\varepsilon{\Theta}_{12}^T\varepsilon{\Theta}_{13}\,\varepsilon \cr
       g_2\,\varepsilon{\Theta}_{13}^T\varepsilon-\varepsilon{\Theta}_{23}^T\varepsilon{\Theta}_{12}^T\,\varepsilon
~&~  g_1\,\varepsilon{\Theta}_{23}^T\varepsilon+\varepsilon{\Theta}_{13}^T\varepsilon{\Theta}_{12}\,\varepsilon
~&~  ({b}_{12}^2-g_1\,g_2)\,\varepsilon
\end{array}\right)~.    
\ee
Thus, the general condition for $\wt{f}$ have inverse is that its Pfaffian does not vanish, or equivalently, $b^2\neq{g_1\,g_2\,g_3}$, on the contrary, the symplectic matrix is singular and, consequently, the system presents a symmetry. It follows some identities:
\ba
&&\Theta_{ij}\,\varepsilon\,\Theta_{ij}^T=\Theta_{ij}^T\,\varepsilon\,\Theta_{ij}=b_{ij}^2\,\varepsilon  ~,  \\
&& ~ \nonumber \\
&&\Theta_{13}\,\varepsilon\,\Theta_{23}^T\,\varepsilon\,\Theta_{12}^T\,\varepsilon 
-\Theta_{12}\,\varepsilon\,\Theta_{23}\,\varepsilon\,\Theta_{13}^T\,\varepsilon=\Tr(\Theta_{13}\,\varepsilon\,\Theta_{23}^T\,\varepsilon\,\Theta_{12}^T\,\varepsilon )\,\I~.
\ea


\begin{thebibliography}{99}

\bibitem{ST} A. Connes, M. R. Douglas, and, A. S. Schwarz, ``Noncommutative geometry and matrix theory: Compactification on tori,'' {J. High Energy Phys.} \textbf{02}, 003 (1998).
\bibitem{STA}  M. R. Douglas and  C. M. Hull, ``D-branes and the noncommutative torus,'' {J. High Energy Phys.} \textbf{02}, 008 (1998).
\bibitem{snyder} S. H. Snyder, ``The Electromagnetic Field in Quantized Space-Time,'' {Phys. Rev.} \textbf{71}, 38--41 (1947).
\bibitem{ST1} M. M. Sheikh-Jabbari, ``SuperYang-Mills theory on noncommutative torus from open strings interactions,'' {Phys. Lett. B} \textbf{450}, 119--125 (1999).
\bibitem{ST1A} N. Seiberg and E. Witten, ``String theory and noncommutative geometry,'' {J. High Energy Phys.} \textbf{09}, 032 (1999).
\bibitem{JC} R. Jackiw, ``Physical instances of noncommuting coordinates,'' {Nucl. Phys. Proc. Suppl.} \textbf{108}, 30--36  (2002).
\bibitem{JC1} S. Carroll, G. B. Field, and R. Jackiw, ``Limits on a Lorentz and Parity Violating Modification of Electrodynamics,'' {Phys. Rev. D} \textbf{41}, 1231--1240 (1990).
\bibitem{JC2} J. Harvey and S. Naculich, ``Cosmic Strings From Pseudoanomalous U(1)\textit{s},'' {Phys. Lett. B} \textbf{217}, 231--237 (1989).
\bibitem{dirac}  P. A. M. Dirac, ``The fundamental equations of quantum mechanics,'' {Proc. Roy. Soc. A} \textbf{109}, 642--653  (1925).
\bibitem{dirac1} P. A. M. Dirac, ``On quantum algebras,'' {Proc. Cambridge Phil. Soc.} \textbf{23}, 412--418  (1926).
\bibitem{peierls}  R. Peierls, Z. Phys. \textbf{80}, 763 (1933)
\bibitem{heisenberg1} W. Heisenberg, ``\"Uber den anschaulichen inhalt der quantentheoretischen kinematik und mechanik,'' Z. Phys. 43, 172–198 (1927) [``The physical content of quantum kinematics and mechanics,'' in Quantum Theory and Measurement, edited by J. A. Wheeler and W. H. Zurek (Princeton University Press, Princeton, 1983) (in English)].
\bibitem{heisenberg2} W. Heisenberg, ``The Physical Principles of the Quantum Theory,'' (Univ. of Chicago Press, Chicago, 1930; Dover,
New York, 1949, 1967).
\bibitem{HG} H. Groenewold, ``On the Principles of elementary quantum mechanics,'' {Physica} \textbf{12}, 405--460  (1946).
\bibitem{MOYAL} J. E. Moyal, ``Quantum mechanics as a statistical theory,'' {Proc. Cambridge Phil. Soc.} \textbf{45}, 99--124 (1949).
\bibitem{wigner} E. Wigner, ``On the quantum correction for thermodynamic equilibrium,'' {Phys. Rev.} \textbf{40}, 749--759 (1932).
\bibitem{flato} F. Bayen, M. Flato, C. Fronsdal, A. Lichnerowicz, and D. Sternheimer, ``Deformation Theory and Quantization. 1. Deformations of Symplectic Structures,'' {Ann. Phys.} \textbf{111}, 61--110; ibid. 111--151 (1978).
\bibitem{fedosov} B. V. Fedosov, ``A Simple geometrical construction of deformation quantization,'' {J. Diff. Geom.} \textbf{40}, 213--238 (1994).
\bibitem{kont}  M. Kontsevich, ``Deformation quantization of Poisson manifolds. 1,'' {Lett. Math. Phys.} \textbf{66}, 157--216 (2003).
\bibitem{QMPS1} D. Fairlie, ``The Formulation of quantum mechanics in terms of phase space functions,'' {Proc. Cambridge Phil. Soc.} \textbf{60}, 581--586  (1964).
\bibitem{QMPS2} C. Zachos, D. Dairlie, and T. Curtright, ``Quantum Mechanics in Phase Space" (World Scientific, 2005).
\bibitem{QMPS3} J. M. Carmona, J. L. Cortes, J. Gamboa, and F. Mendez, ``Noncommutativity in field space and Lorentz invariance violation,'' {Phys. Lett. B} \textbf{565}, 222--228 (2003).
\bibitem{QMPS4} J. M. Carmona, J. L. Cortes, J. Gamboa, and F. Mendez, ``Quantum theory of noncommutative fields,'' {J. High Energy Phys.} \textbf{0303}, 058 (2003).
\bibitem{ANO2}  E. M. C. Abreu, C. Neves, and W. Oliveira, ``Noncommutativity from the symplectic point of view ,'' {Int. J. Mod. Phys. A} \textbf{21}, 5359 (2006).
\bibitem{FJ2} L. D. Faddeev and R. Jackiw, ``Hamiltonian Reduction of Unconstrained and Constrained Systems,'' {Phys. Rev. Lett.} \textbf{60}, 1692--1694 (1988).
\bibitem{FJ3} N. M. J. Woodhouse,  ``Geometric Quantization,'' (Clarendon Press, Oxford, 1980).
\bibitem{FJ4} J. Barcelos-Neto and C. Wotzasek, ``Symplectic quantization of constrained systems,'' {Mod. Phys. Lett. A} \textbf{7}, 1737--1747 (1992).
\bibitem{FJ5} J. Barcelos-Neto and C. Wotzasek, ``Faddeev-Jackiw quantization and constraints,'' {Int. J. Mod. Phys. A} \textbf{7}, 4981--5003  (1992).
\bibitem{ANO1} J. A. Neto, C. Neves, and W. Oliveira, ``Gauging the SU(2) Skyrme model,'' {Phys. Rev. D} \textbf{63}, 085018  (2001).
\bibitem{ANO1a} A. C. R. Mendes, C. Neves, W. Oliveira, and D. C. Rodrigues, ``Symplectic embedding of second class systems,'' {Nucl. Phys. B (Proc. Suppl.)} \textbf{127}, 170--173 (2004).
\bibitem{ANO1b} A. C. R. Mendes, C. Neves, W. Oliveira, and F. I.Takakura, ``Symplectic embedding of a fluid dynamical model,'' {J. Phys. A: Math. Gen.} \textbf{37}, 1927--1944 (2004).
\bibitem{ANO1c} E. M. C. Abreu, J. A. Neto, A. C. R. Mendes, C. Neves, and W. Oliveira, ``Obtaining gauge invariant actions via symplectic embedding formalism,'' {Annalen Phys.} \textbf{524}, 434--455 (2012).
\bibitem{WOCN} C. Neves and W. Oliveira, ``Clebsch parametrization from the symplectic point of view,'' {Phys. Lett. A} \textbf{321}, 267--272 (2004).
\bibitem{ANO2a} E. M. C. Abreu, J. A. Neto, A. C. R. Mendes, C. Neves, W. Oliveira, and M. V. Marcial, ``Lagrangian formulation for noncommutative nonlinear systems ,'' {Int. J. Mod. Phys. A} \textbf{27}, 1250053 (2012).
\bibitem{ANO2b} E. M. C. Abreu, A. C. R. Mendes, and W. Oliveira, ``Noncommutativity and Duality through the Symplectic Embedding Formalism,'' {SIGMA} \textbf{6}, 059 (2010).
\bibitem{mezincescu} L. Mezincescu, ``Star operation in quantum mechanics,'' e-print arXiv:hep-th/0007046v2.
\bibitem{bopp1} T. Curtright, D. Fairlie, and  C. K. Zachos, ``Features of time independent Wigner functions,'' {Phys. Rev. D} \textbf{58}, 025002 (1998).
\bibitem{bopp2} J. Gamboa, M. Loewe, and J. C. Rojas, ``Noncommutative quantum mechanics,'' {Phys. Rev. D} \textbf{64}, 067901  (2001).
\bibitem{djemai1} A. E. F. Djemai, `` On noncommutative classical mechanics,'' Int. J. Theor. Phys. \textbf{43} 299 (2004).
\bibitem{djemai2} A. E. F. Djemai, ``On quantum mechanics on noncommutative quantum phase space,'' Commun. Theor. Phys. \textbf{41}, 837-844  (2004).
\bibitem{gosson2} N. Dias, M. de Gosson, F. Luef, and J. Prata, ``A Deformation Quantization Theory for Non-Commutative Quantum Mechanics,'' J. Math. Phys. \textbf{51} (2010).
\bibitem{gosson3} M. de Gosson, ``A pseudodifferential calculus on non-standard symplectic space,'' Appl. Anal. \textbf{90}(11), 1665--1676, (2011).
\bibitem{gosson1} M. de Gosson and F. Luef, ``Born--Jordan Pseudodifferential Calculus, Bopp Operators and Deformation Quantization,'' Integr. Equations and Oper. Theory \textbf{84}(4), 463--485 (2016).
\bibitem{Jackiw}  R. Jackiw, ``(Constrained) Quantization Without Tears,'' e-print arXiv:hep-th/9306075v1, Published in ``R. Jackiw: Diverse topics in theoretical and mathematical physics'' 367-381 (1993) 
\bibitem{chaichian} M. Chaichian, M. M. Sheikh-Jabbari, and A. Tureanu, ``Hydrogen atom spectrum and the Lamb shift in noncommutative QED,'' Physical Review Letters \textbf{86} (13), 2716 (2001)
\bibitem{BG} R. Banerjee and S. Ghosh, ``The Chiral oscillator and its applications in quantum theory,'' {J. Phys. A: Math. Gen.} \textbf{31}, {L603}--{L608} (1998).
\bibitem{Lewis1} H. R. Lewis, W. E. Lawrence, and J. D. Harris, ``Quantum Action-Angle Variables for the Harmonic Oscillator,'' {Phys. Rev. Lett.} \textbf{77}, 5157--5159 (1996).
\bibitem{Lewis2} H. R. Lewis and W. B. Riesenfeld, ``An Exact quantum theory of the time dependent harmonic oscillator and of a charged particle time dependent electromagnetic field,'' {J. Math. Phys.} \textbf{10}, 1458--1473 (1969).
\bibitem{BS} D. Bigatti and L. Susskind, ``Review of matrix theory,'' e-print arXiv:hep-th/9712072, Published in ``Cargese 1997, Strings, branes and dualities'' 277-318 (1997)
\bibitem{BFFT0} I. A. Batalin and E. S. Fradkin, ``Operator Quantization of Dynamical Systems With Irreducible First and Second Class Constraints,'' Phys. Lett. B \textbf{180}, 157 (1986).
\bibitem{BFFT1} I. A. Batalin and E. S. Fradkin, ``Operatorial Quantization of Dynamical Systems Subject to Second Class Constraints,'' Nucl. Phys. B \textbf{279}, 514 (1987).
\bibitem{BFFT1a} I. A. Batalin, E. S. Fradkin, and T. E. Fradkina, ``Another Version for Operatorial Quantization of Dynamical Systems With Irreducible Constraints,'' Nucl. Phys. B \textbf{314} (1989) 158-174, Erratum: Nucl. Phys. B \textbf{323}, 734-735 (1989). 
\bibitem{BFFT2} I. A. Batalin and I. V. Tyutin, ``Existence theorem for the effective gauge algebra in the generalized canonical formalism with Abelian conversion of second class constraints,'' Intl. J. Mod. Phys. A \textbf{6}, 3255 (1991).
\bibitem{FS} L. D. Faddeev and S. L. Shatashvili, ``Realization of the Schwinger Term in the Gauss Law and the Possibility of Correct Quantization of a Theory with Anomalies,'' Phys. Lett. B \textbf{167}, 225 (1986).
\bibitem{CW} C. Wotzasek, ``On the {Wess-Zumino} Term for a General Anomalous Gauge Theory With Second Class Constraints,'' Int. J. Mod. Phys. A \textbf{5}, 1123 (1990).
\end{thebibliography}
\end{document}